\documentclass{ws-procs9x6}

\begin{document}
\newcommand{\ome}{\omega_{\rm rot}}
\newcommand{\Su}{${}^{32}$S}
\newcommand{\Ar}{${}^{36}$Ar}
\newcommand{\Ca}{${}^{40}$Ca}
\newcommand{\Ti}{${}^{44}$Ti}
\newcommand{\Cr}{${}^{48}$Cr}
\newcommand{\NZ}{$N$=$Z$~}

\title{
Mean-field approach to superdeformed high-spin states \\
in $^{40}$Ca and neutron-rich $^{50}$S regions
}

\author{
T.~Inakura, M.~Yamagami, and K. Matsuyanagi
}
\address{
Department of Physics, Graduate School of Science,\\
Kyoto University, Kitashirakawa, Kyoto 606-8502, Japan
}  

\author{
S.~Mizutori
}
\address{
Department of Human Science, Kansai Women's College,\\
Kashiwara City, Osaka 582-0026, Japan
}  

\maketitle

\abstracts{
With the use of the symmetry-unrestricted cranked SHF
method in the 3D coordinate-mesh representation,
a systematic search for the
SD and HD rotational bands in the \NZ nuclei from \Su~  to \Cr~ 
has been done, and
SD and HD solutions have been found in ~\Su, \Ar, \Ca, \Ti,~ 
and in ~\Ar, \Ca, \Ti, \Cr, respectively.
The SD band in ~\Ca~ is found to be extremely soft against 
both the axially symmetric ($Y_{30}$) and 
asymmetric ($Y_{31}$) octupole deformations.
Possible presense of SD states in neutron-rich sulfur isotopes
from $^{46}$S to $^{52}$S has also been investigated, and
deformation properties of
neutron skins both in the ground and SD states are discussed. 
}

\section{Introduction}

Quite recently, superdeformed(SD) rotational bands
have been discovered in \Ar, \Ca, and \Ti.\cite{sve00,ide01,lea00}
One of the important new features of them is that they
are built on excited $0^+$ states and observed up to high spin,
in contrast to the SD bands in heavier mass regions
where low-spin portions of them are unknown in almost all cases.
In this talk, we shall first report results of the symmetry-unrestricted,
cranked Skyrme-Hartree-Fock (SHF) calculations for
these SD bands.
The calculation has been carried out 
with the use of the fully three-dimensional (3D), Cartesian coordinate-mesh 
representation without imposing any symmetry restriction.
\cite{yam00,yam01,ina02}
The computational algorithm is basically the same 
as in the standard one \cite{bon87}
except that the symmetry restrictions are removed.
For comparison sake, 
we also carry out the standard symmetry-restricted calculations 
imposing reflection symmetries about 
the $(x,y)$-, $(y,z)$- and $(z,x)$-planes. 
By comparing these results, we can clearly identify effects of 
reflection symmetry breaking in the mean field.

We shall next present results of the cranked SHF calculation for  
SD bands in the neutron-rich sulfur isotopes near the neutron 
drip line. These nuclei are expected to constitute a new 
``SD doubly closed'' region associated with the
SD magic numbers, $Z=16$ for protons and $N \simeq 30$-32 for neutrons. 
An interesting theoretical subject for the SD bands in these 
neutron-rich region is to understand deformation properties of 
neutron skins.
We shall discuss on this point.

\section{\Ca~region}


\begin{figure}[ht]
\centerline{\epsfxsize=2.75in\epsfbox{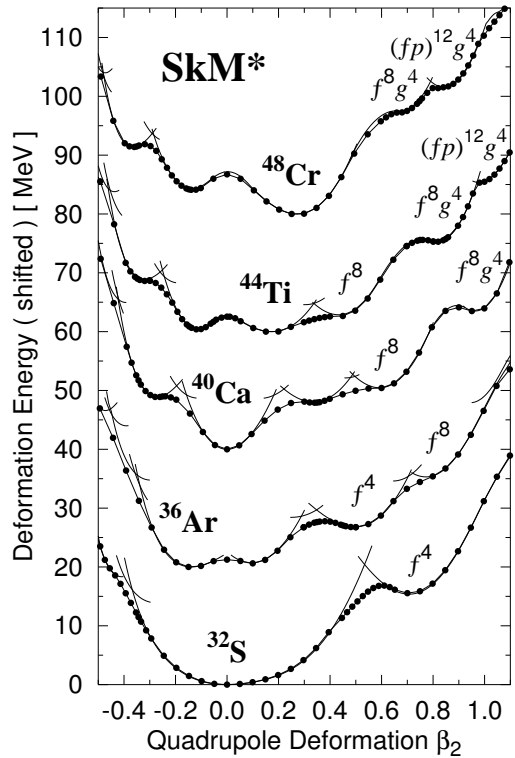}}
\caption{
Deformation energy curves as functions of the quadrupole deformation 
$\beta_2$ calculated at $I=0$ by means of the constrained SHF procedure 
with the SkM$^*$ interaction. 
The axial-asymmetry parameter $\gamma$ is constrained to be zero.
The curves for different nuclei are shifted 
by 20 MeV in order to facilitate the comparison.
Solid lines with and without filled circles represent 
the results obtained by the
unrestricted and restricted versions, respectively (see the text).
The notations $f^n g^m$ and $(fp)^n g^m$ indicate 
the configurations in which 
the $f_{7/2}$ shell ($fp$ shell) and the $g_{9/2}$ shell
are respectively occupied by $n$ and $m$ nucleons.}
\end{figure}

\begin{figure}[ht]
\epsfxsize=10cm   
\centerline{\epsfxsize=3.2in\epsfbox{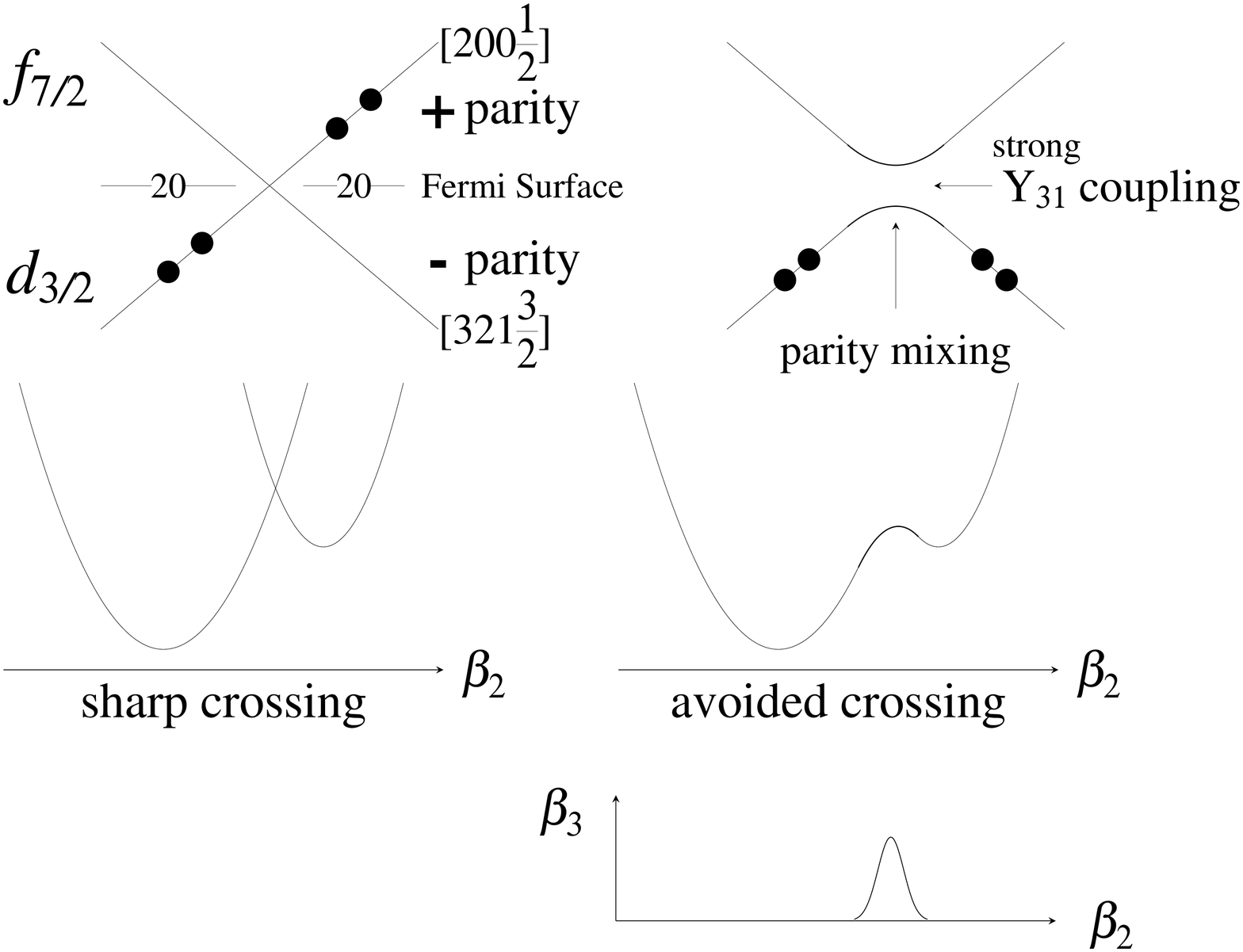}}   
\caption{
Level crossing between single-particle levels having opposite parities.
Configuration rearrangement can take place by breaking the reflection symmetry
in the mean field. The $r^3 Y_{31}$ matrix element between levels with the 
asymptotic quantum numbers $\left[321 \frac 32 \right]$ and 
$\left[200 \frac 12 \right]$ is large,
and significantly contributes to the tendency toward the non-axial
$Y_{31}$ octupole deformation discussed in the text.}
\end{figure}

\begin{figure}[ht]
\centerline{\epsfxsize=3.25in\epsfbox{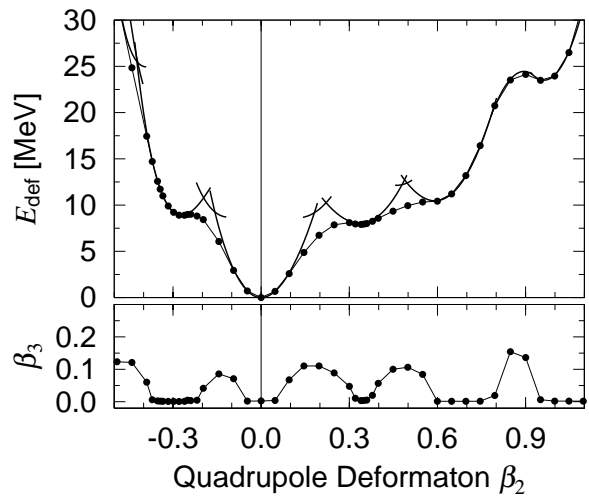}}   
\caption{
The upper part: Deformation energy curve for ~\Ca~(same as in Fig.~1).
The lower part: Octupole deformation $\beta_3$ 
obtained by the unrestricted SHF calculation with SkM$^*$,
plotted as a function of $\beta_2$.}
\end{figure}

\begin{figure}[ht]
\centerline{\epsfxsize=3.9in\epsfbox{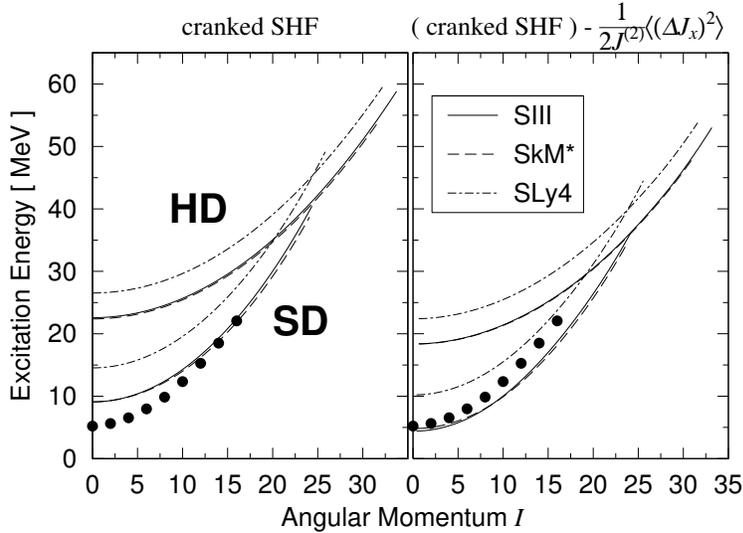}}
\caption{
Comparison between the excitation energies of the SD and HD bands 
in ~\Ca, calculated by using different versions 
of the Skyrme interaction and 
the experimental data (filled circles). 
Solid, dashed and dashed-dotted lines indicate
the results with the SIII, SkM$^*$ and SLy4 interactions, respectively.
Results with and without including the zero-point rotational energy 
correction are shown in the right- and left-hand sides, respectively.}
\end{figure}

\begin{figure}[ht]
\centerline{\epsfxsize=3.0in\epsfbox{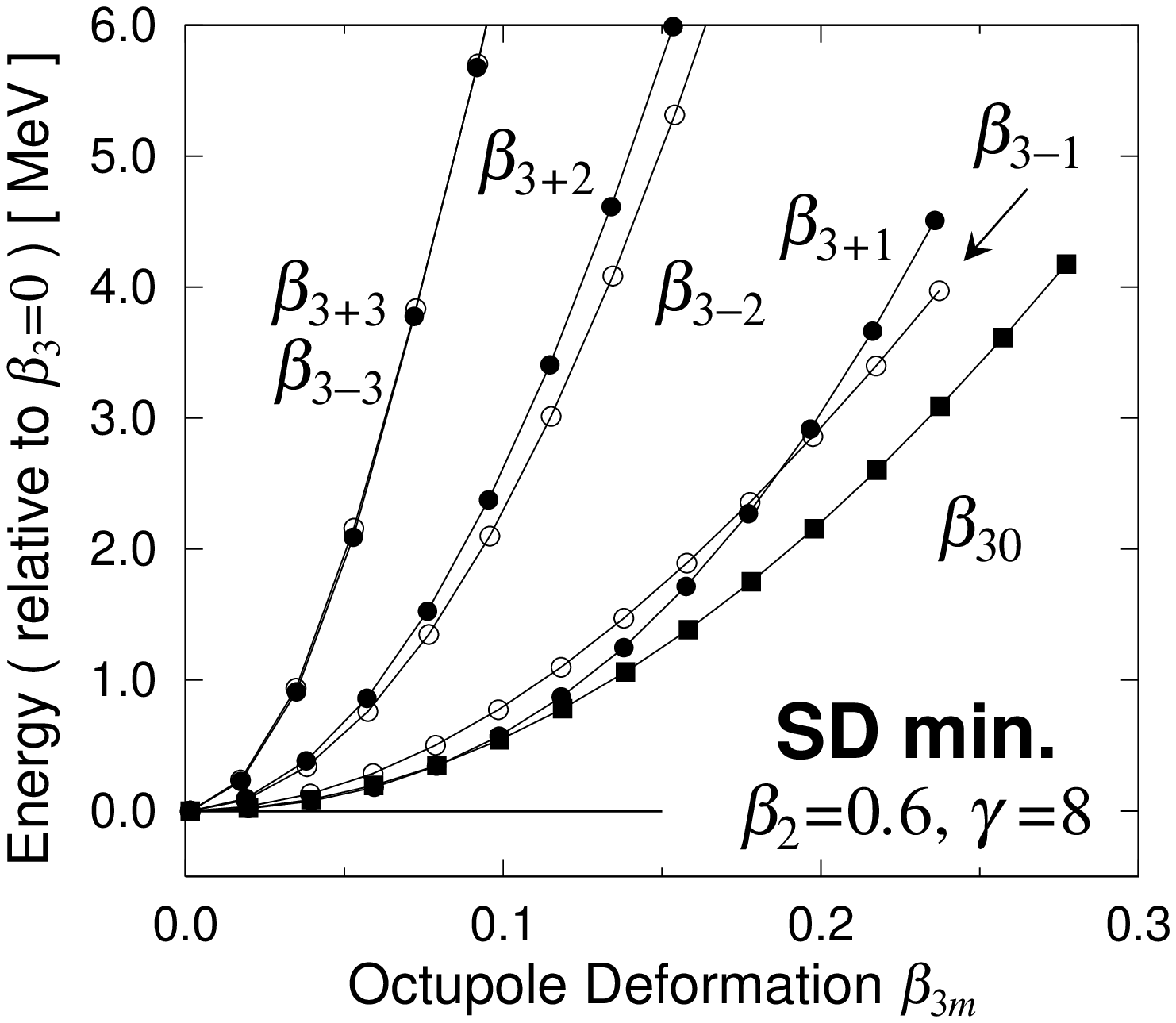}}   
\caption{
Deformation energy curves 
(measured from the energy at $\beta_3=0$) as functions of 
the octupole deformation parameters  $\beta_{3m}(|m|=0,1,2,3)$,  
calculated for ~\Ca~ by means of the constrained HF procedure
with SkM$^*$.
The quadrupole deformation parameters are fixed 
at $\beta_2=0.6$ and $\gamma=8^{\circ}$. 
One of the $\beta_{3m}(|m|=0,1,2,3)$ is varied while the other
$\beta_{3m}$'s are fixed to zero.}
\end{figure}


Figure 1 shows deformation energy curves   
evaluated by means of the constrained HF procedure.
Solid lines with and without filled circles in these figures 
represent results 
of unrestricted and restricted calculations, respectively.
In both cases, 
we otain local minima corresponding to the SD states
for \Su, \Ar, \Ca~ and \Ti~ 
in the region $0.4 \le \beta_2 \le 0.8$.
(The local minimum in \Ti~ is triaxial so that it is not clearly seen 
the $\gamma=0$ section.)
The local minima in \Su~ and \Ar~ involve four particles
in the $fp$ shell, while those in
\Ca~ and \Ti~ involve eight particles.
In addition to these SD minima, 
we also obtain local minima 
in the region $\beta_2 \geq 0.8$ for \Ca, \Ti~ and \Cr. 
These minima involve additional four particles 
in the single-particle levels that reduce to the g$_{9/2}$ levels
in the spherical limit.
Somewhat loosely we call these local minima ``hyperdeformed (HD)."
The HD solution in ~\Ca~ corresponds to the 12p-12h configuration.

We notice in this figure that the crossings between configurations involving
different numbers of particles in the $fp$ shell are sharp in the
restricted calculations, while we always obtain smooth configuration 
rearrangements in the unrestricted calculations.
The reason for this different behavior between the unrestricted and
restricted calculations is rather easy to understand:
When the parity symmetry is imposed, there is no way, within the mean-field
approximation, to mix configurations having different number of particles 
in the $fp$ shell.
In contrast, as illustrated in Fig.~2, smooth crossover between these 
different configurations is possible via mixing between positive- and 
negative-parity single-particle levels, 
when such a symmetry restriction is removed.   
Octupole deformation parameters $\beta_3$ 
are plotted as functions of $\beta_2$ in the lower 
portion of Fig.~3 for the case of \Ca.
We see that values of $\beta_3$ are zero near the local minima,
but rise in the crossing region. 
This means that the configuration
rearrangements in fact take place through paths in the deformation space 
that break the reflection symmetry.

Excitation energies of the SD and HD bands in \Ca~ 
calculated by using different versions (SIII, SkM$^*$, SLy4)
of the Skyrme interaction are compared  
with the experimental data \cite{ide01}
in the left-hand portion of Fig.~4.
The SD band is slightly triaxial 
with $\gamma = 6^{\circ}$-$9^{\circ} (8^{\circ}$-$9^{\circ})$
and it terminates at $I \simeq 24$ 
for the SIII (SkM$^*$) interaction.
(In the case of \Ti, the shape is more triaxial 
with $\gamma = 18^{\circ}$-$25^{\circ}$ and
$13^{\circ}$-$19^{\circ}$,
and the SD band terminates at $I \simeq 12$ and 16
for the SIII and SkM$^*$ interactions, respectively.\cite{ina02})
One may notice that the excitation energy of 
the SD band-head state is overestimated. 
We have evaluated the zero-point rotational energy correction,
$\frac{1}{2J^{(2)}}\langle(\Delta \hat{J}_x)^2\rangle$, 
as a function of the rotational frequency $\ome$.
Here, $\Delta \hat{J}_x = \hat{J}_x - \langle\hat{J}_x\rangle$ and
$J^{(2)}$ denotes the dynamical moment of inertia defined by
$J^{(2)}=dI/d\ome$.
Excitation energies including this correction are shown 
in the right-hand portion of Fig.~4. 
We see that the calculated excitation energies are
significantly improved by including this correction.

Let us examine stabilities of the SD local minimum in \Ca~ 
against octupole deformations.
Figure 5 shows deformation energy curves as functions of 
the octupole deformation parameters  $\beta_{3m}(|m|=0,1,2,3)$  
about the SD minimum.
We immediately notice that the SD state is extremely soft 
with respect to the $\beta_{30}$ and $\beta_{31}$ deformations.

Quite recently, Imagawa and Hashimoto have carried out
a selfconsistent RPA calculation in the 3D Cartesian-mesh representation
on the basis of the SHF mean field,
and they have obtained, for the SIII (SkM$^*$) interaction,
a strongly collective octupole vibrational mode with $K^\pi=1^-$ 
at about 1.1 (0.6) MeV excitation from the SD band head.\cite{ima02}
Thus, it is extremely interesting to search for negative-parity
rotational bands associated with the non-axial $K^\pi=1^-$ 
octupole vibrational modes built on the SD yrast band.

\section{$^{50}$S~region}

Figure 6 shows deformation energy curves   
for neutron-rich sulfur isotopes from $^{46}$S to $^{52}$S,
which indicates that the SD local minima is deepest at $^{50}$S.
As shown in Fig.~7, this result is common for the SHF calculations 
with the use of SIII, SkM$^*$ and SLy4 interactions.
Thus, the neutron SD shell structure seems to be slightly modified from 
that known in the Zn region with $N \simeq Z$, where  
$N \simeq 30$-32 are the SD magic numbers.


\begin{figure}[ht]
\centerline{\epsfxsize=2.75in\epsfbox{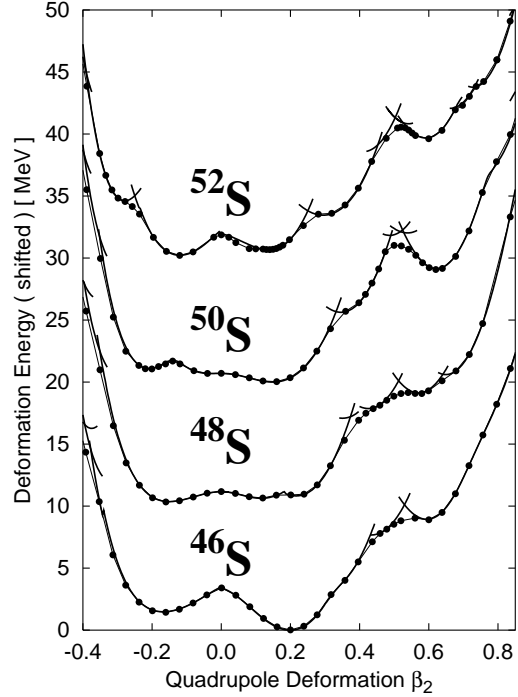}}   
\caption{
Deformation energy curves as functions of $\beta_2$ 
calculated at $I=0$ by means of the constrained SHF procedure with
the SLy4 interaction. 
The axial-asymmetry parameter $\gamma$ is constrained to be zero.
The curves for different nuclei are shifted 
by 10 MeV to facilitate the comparison.
Solid lines with and without filled circles represent 
the results obtained by the
unrestricted and restricted versions, respectively.
The notations $f^n g^m$ indicate 
the configurations in which 
the $f_{7/2}$ shell and the $g_{9/2}$ shell
are respectively occupied by $n$ and $m$ nucleons.}
\end{figure}

\begin{figure}[ht]
\centerline{\epsfxsize=2.5in\epsfbox{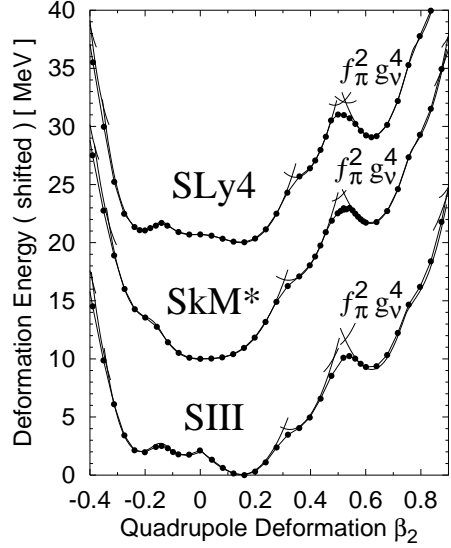}}   
\caption{
Comparison of deformation energy curves for $^{50}$S    
obtained by using different versions of the Skyrme interaction. 
The curves for different versions are shifted 
by 10 MeV.}
\end{figure}

\begin{figure}[ht]
\centerline{\epsfxsize=3.0in\epsfbox{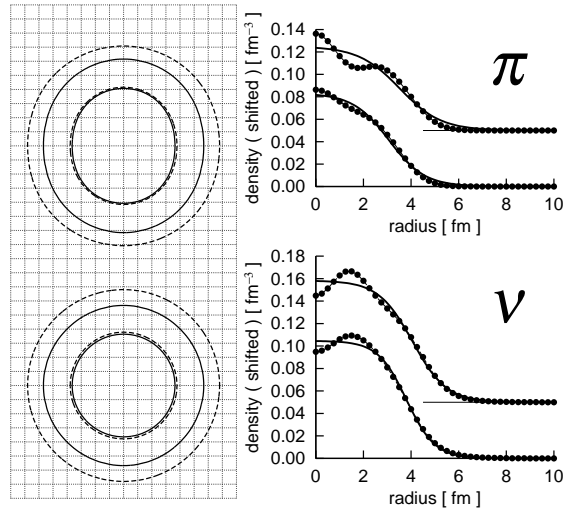}}
\caption{
Neutron and proton density distributions in the ground state of 
$^{50}$S, calculated by the unrestricted SHF with SLy4 and with
a mesh size of 0.25 fm.
Left-hand side: equi-density lines with $50\%$ and $1\%$ of the central
density in the $(x,z)$ and $(x,y)$ planes are drawn.
Solid and dashed lines indicate those for protons and neutrons, respectively.
Right-hand side: Density distributions along the major and minor axes
are drawn by thin-solid lines with filled circles.
The least-square fits of them with the Wood-Saxon function
are also shown by solid lines. The former densities are shifted up
by 0.05 fm$^{-3}$ to facilitate the comparison.}
\end{figure}

\begin{figure}[ht]
\centerline{\epsfxsize=3.0in\epsfbox{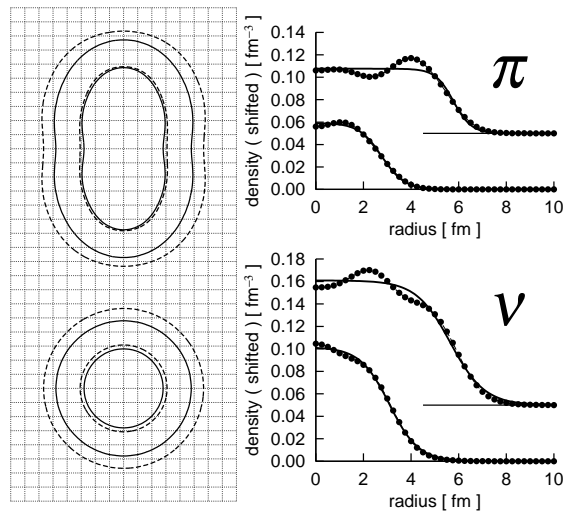}}   
\caption{
The same as Fig.~8 but for the SD state.}
\end{figure}

\begin{table}
\tbl{
Root-mean square radii ( $R_{{\rm rms}}$ ), quadrupole deformation
parameters ( $\beta_2$ ), 
half-density radii for the major and minor axes ( $R^z_{1/2}$ ,
$R^x_{1/2}$ ), surface diffuseness parameters along
the major and minor axes 
( $a^z$ , $a^x$ ), 
evaluated
by the SHF method with SLy4  for the SD state 
in $^{50}$S. The proton and neutron contributions are separately
shown together with their differences.
\vspace*{1pt}}
{\footnotesize
\tabcolsep7pt
\begin{tabular}{|c|c|c|c|c|c|c|}
\hline
{} &{} &{} &{} &{} &{} &{}\\[-1.5ex]
{SD} & $R_{\rm rms}$ & $\beta_2$ & $R^z_{1/2}$ & $R^x_{1/2}$ & $a^z$ & $a^x$ \\[1ex]
\hline
{} &{} &{} &{} &{} &{} &{}\\[-1.5ex]
neutrons    & 4.15 & 0.59 &  5.65 & 3.09 & 0.72 & 0.61 \\[1ex]
protons     & 3.75 & 0.70 &  5.76 & 2.77 & 0.38 & 0.47 \\[1ex]
differences & 0.41 & -0.11 & -0.12 & 0.32 & 0.34 & 0.14 \\[1ex]
\hline

\end{tabular}}
\vspace*{-13pt}
\end{table}


Figures 8 and 9 show the neutron and proton density profiles for
the ground and the SD states, respectively.
We see that deformed neutron skins are present in both cases.
These calculations are done with use of a small mesh size of 0.25 fm.
In order to examine deformation properties of these neutron skins
in more detail, we have made a least square fitting to the density
distribution along each principal axis direction
with the Woods-Saxon function. The half-density radii and 
the diffuseness parameters extracted in this way for the SD state are listed in Table 1.
We see that the neutron skin is formed mainly due to the difference in the
diffuseness between protons and neutrons (rather than the difference in
the half-density radius). It is interesting to note that the proton 
diffuseness parameter along the major axis is significantly smaller 
than that along the minor axis.

The presence of the deformed neutron skins may be detected through 
excitation spectra of these nuclei. 
Thus, search for soft $K^\pi=0^-$ and $1^-$
 (dipole + octupole) vibrational modes
in unstable nuclei with deformed neutron skins
seems especially interesting.
Note that octupole modes will be mixed with dipole modes
in deformed nuclei.
For studying these modes, we need to develop the SHF-Bogoliubov + quasiparticle RPA approach
such that we can take into account deformation, pairing, and continuum
effects simultaneously.
We can further envisage to go
beyond the  quasiparticle RPA by means of the
selfconsistent collective coordinate method.\cite{mar80}

\clearpage
\section{Summary}

With the use of the symmetry-unrestricted cranked SHF
method in the 3D coordinate-mesh representation,
we have carried out a systematic theoretical search for the
SD and HD rotational bands in the \NZ nuclei from \Su~ to \Cr.
We have found the SD solutions in \Su, \Ar, \Ca, \Ti,~and 
the HD solutions in \Ar, \Ca, \Ti, \Cr.
Particular attention has been paid to the recently discovered SD
band in \Ca, and 
we have found that the SD band in \Ca~is extremely soft against 
both the axially symmetric ($Y_{30}$) and 
asymmetric ($Y_{31}$) octupole deformations.
Thus, it will be especially interesting to search for negative-parity
rotational bands associated with non-axial $K^\pi=1^-$ 
octupole vibrations  built on the SD yrast band.

We have also discussed possible presense of SD states in sulfur isotopes
from $^{46}$S to $^{52}$S, which are situated near the neutron drip line.
An interesting subject in this region is the appearance of deformed
neutron skins both in the ground and SD states. 
The presence of the deformed neutron skins may be detected through 
excitation spectra of these nuclei. 
Thus, search for new kinds of soft $K^\pi=0^-$ and $1^-$
 (dipole + octupole) vibrational modes of excitation is challenging,
both theoretically and experimentally.

\end{document}